\documentclass[prb,aps,nofootinbib,superscriptaddress,floatfix,10 pt,twocolumn,longbibliography]{revtex4-2}
\usepackage[urlcolor=blue,colorlinks=true,citecolor=blue,linkcolor=blue,bookmarks=false]{hyperref}
\usepackage{amsmath,graphicx, color}
\usepackage{float}
\usepackage{dcolumn}
\newcolumntype{.}{D{.}{.}{-1}}
\usepackage{amsfonts}
\usepackage[normalem]{ulem}
\newcommand{\neel}{Néel }
\usepackage{orcidlink}
\usepackage{setspace}
\begin{document}
\title{Magnetotransport and electronic band structure of EuNi$_2$As$_2$ antiferromagnet}
\author{Faheem Gul\orcidlink{0000-0002-9705-9900}}
\author{Mane Sahakyan\orcidlink{0000-0002-8867-1356}}
\author{Orest Pavlosiuk\orcidlink{0000-0001-5210-2664}}
%\author{Dariusz Kaczorowski\orcidlink{0000-0002-8513-7422}}
\author{Piotr Wiśniewski\orcidlink{0000-0002-6741-2793}}\email[e-mail: ]{p.wisniewski@intibs.pl}
\affiliation{Institute of Low Temperature and Structure Research, Polish Academy of Sciences, Wrocław, Poland}
%	\pacs{}
\date {\today}
\begin{abstract} 
We investigated the magnetotransport properties of single-crystals of tetragonal van der Waals compound EuNi$_2$As$_2$, that orders antiferromagnetically below 14.6\,K in an incommensurate helical structure. 
Metamagnetic transitions are revealed by the magnetization measured in the magnetic field applied transverse to the axis of the helix, and are clearly reflected in the magnetoresistance. Overall, the magnetoresistance is small, but shows complex changes with the temperature, the strength, and the angle of the applied magnetic field. In magnetically ordered state, magnetoresistance shows prominent anomalies related to the metamagnetic transitions. For temperatures above the \neel point the negative magnetoresistance can be modeled very well  with de Gennes-Friedel mechanism of the spin-disorder-scattering reduction. 
Hall resistivity data indicate hole-dominated multi-band conductivity in antiferromagnetic state and single-band one above the \neel temperature, with carrier concentrations of the order of 10$^{22}\rm cm^{-3}$. This metallic character of the compound seems to obscure the plausible topological contribution to the Hall resistivity. Our \textit{ab-initio} calculations of electronic band structure showed that the electronic structure changes very strongly upon magnetic ordering, but the density of states at the Fermi level differs by a factor smaller than two, in agreement with experimental Hall resistivity data. Meaningful changes in the density of states, magnetic moments, and screening length of Eu-4\textit{f} orbitals are discussed in terms of the effects of Hubbard corrections.
	\end{abstract} 
	\maketitle
\section{Introduction}\label{Intro}
Research on the complex interplay of magnetism and electronic band structure has a long history, but has been strongly rejuvenated with the discovery of topological insulators and semimetals \cite{Armitage2018}. 
Eu-based compounds provide fertile ground for this endeavor, mainly due to the strong magnetic moments of Eu$^{2+}$ ion and lack of a crystalline-field effect \cite{Chen2025rev}. Recent studies of several Eu-based systems of 1:2:2 stoichiometry, such as EuCd$_2$As$_2$ \cite{Rahn2018,Ma2019}, EuCd$_2$Sb$_2$ \cite{Soh2018,gul2026origin},  %EuZn$_2$As$_2$ \cite{Yi2023},  
EuZn$_2$Sb$_2$ \cite{Singh2024}, EuIn$_2$As$_2$ \cite{Yan2022}, revealed that the magnetic structure can induce or significantly alter topologically non-trivial electronic band structure. 

Interestingly, all these compounds exhibit a pronounced topological Hall effect (THE). In EuCd$_2$As$_2$ and EuCd$_2$Sb$_2$, both Weyl nodes and scalar chirality of non-coplanar spins induce a Berry curvature that gives rise to THE. In EuZn$_2$Sb$_2$ and EuIn$_2$As$_2$ solely the second mechanism plays a role, although with various backgrounds: in the former compound, the spin chirality occurs within the domain wall, whereas in the latter, it is brought by the magnetic field applied to the helical magnetic structure \cite{Yan2022, Singh2025}. 

Another europium compound, but from the 1:1:1 family, namely, EuCuAs, brought additional motivation for this work. It also has the incommensurate helical magnetic structure \cite{Soh2024EuCuAs}. Such a magnetic structure breaks an inversion and time reversal symmetry, which induces Weyl nodes leading, in turn, to the rise of the topological Hall effect \cite{Roychowdhury2023, Soh2024EuCuAs}. All the mentioned above compounds exhibit antiferromagnetism due to Ruderman–Kittel–Kasuya–Yosida interactions and they are semimetals with carrier concentrations of the order of magnitude of 10$^{19}$\,cm$^{-3}$. 

Upon looking for other helical antiferromagnets, that might by exhibiting THE, we noticed the EuNi$_2$As$_2$ compound crystallizing in a body-centered tetragonal ThCr$_2$Si$_2$-type structure ($I4/mmm$ space group) \cite{Sangeetha2019, Jin2019}. Resistivity measurement has shown that the system is more metallic than all previously mentioned compounds. It orders antiferromagnetically below the Néel temperature $T_{\rm N}$=\,14.6\,K. 
The neutron powder-diffraction analysis has revealed that it has an incommensurate helical magnetic structure with spins stacked ferromagnetically in the \textit{ab}-planes and arranged helically around the \textit{c}-axis, with the $(0,0,0.92)$ propagation vector \cite{Jin2019}. Due to these interesting features, the study of magnetotransport properties of this compound seemed indispensable. Therefore we grew single crystals of EuNi$_2$As$_2$, and present below the results of performed on them magnetic and magnetotransport measurements. 

There was a recent theoretical report in which, the results of \textit{ab-initio} density-functional theory (DFT), using a simple GGA approximation \cite{tran2025first}, have been presented. They included the electronic and magnetic properties of EuFe$_{2-x}$Ni$_2$As$_2$ solid solution, but with an emphasis on the effect of Ni substitution on the possible emergence of superconductivity in the system.

Here, we also present the results of fully relativistic DFT calculations of the band structure of the compound, but performed in a more extended way  compared to that described in \cite{tran2025first} (focusing on the effects of Hubbard term corrections), including dispersions of energy bands, densities of states, magnetic moments and screening length of Eu-4\textit{f} orbitals.
\section{Methods}
\noindent\textbf{Experimental}

The high-quality single crystals of EuNi$_2$As$_2$ were grown using the Bi flux (for details see Supplemental Material \cite{SupplMater}). 
The energy-dispersive x-ray spectroscopy (EDS) using NanoSEM 230 (FEI Nova) scanning electron microscope equipped with an EDAX Genesis XM4 spectrometer was employed to examine their chemical composition. 

The quality and orientation of the single crystals were determined by Laue back-scattering x-ray diffraction using the LAUE-COS (Proto Mfg.) system.  
Magnetic properties measurements were carried out with the SQUID magnetometer (MPMS-XL, Quantum Design). 
Electrical transport measurements were performed on the PPMS (Quantum Design) platform.
The bar-shaped samples were cut from single crystals for the electrical transport measurements, which we carried out using the standard four-probe technique.
Electrical contacts were prepared with 100\,$\mu$m-thick silver wires attached to the sample with silver paste. The heat capacity was measured on the PPMS platform by the relaxation method.\\\\
\noindent\textbf{Computational details}

We investigated the electronic band structure of EuNi$_2$As$_2$ in the following steps: (i) first, we performed a unit cell optimization on the lattice parameters using the Pseudo Potential Projector Augmented Wave (PP-PAW) approach (VASP software version 5.4.4) \cite{kresse1999ultrasoft}. As starting parameters, we took the lattice parameters reported for EuNi$_2$As$_2$ in literature: $a$=\,0.4101 nm and $c$=\,1.0046 nm \cite{Jin2019}. % c\a=2.45
The optimization was performed with Perdew–Burke–Ernzerhof (PBE) generalized gradient (GGA) exchange–correlation functional \cite{PhysRevLett.100.136406}. The relaxed lattice parameters of EuNi$_2$As$_2$  $a$=\,0.4001 nm, $c$=\,1.0297 nm, are comparable to the previously reported \cite{tran2025first}. Thus, the optimized unit cell volume was reduced by 2.47$\%$.\\ 
(ii) The DFT calculations of electronic band structure of EuNi$_2$As$_2$ were performed by Full-Potential Linearized Augmented Plane-Wave plus local orbitals (FP-LAPW+lo) method  \cite{wimmer1981full}. The  DFT calculations have been done using the ELK computational code (version 8.8.26) \cite{ELK2}, employing the local spin density approximation with the inclusion of Coulomb correlations (LSDA+U) \cite{perdew1992accurate}. %by using the “highq = true” option. 
The tetrahedron method of \textit{k} integration was applied for self-consistent calculations with the
$(10\times10\times4) k$-mesh of the Brillouin zone. The plane wave cut-off was determined such that $R^{MT}\times|G+k|_{max}$=\,9.0, with $R^{MT}$ denotes the average atomic radius and G$_{max}$ represents the maximum wave vector value. Muffin-tin geometrically optimized sphere radii were as follows: $R^{MT}_{\rm Eu}$=\,0.148 nm, $R^{MT}_{\rm Ni}$=\,0.117 nm, and $R^{MT}_{\rm As}$=\,0.117 nm.
%\PW{Was that the energy-optimization of the unit cell by ELK, which brought out these radii, Answer:  no it is geometry optimization, it checks the distance between all pairs of atoms, if the sum of the initial radii for two atoms is greater than the distance between them, Elk scales them down, also ensuring a small gap between the spheres to avoid numerical instability at the boundaries
 %The Elk code automatically determined these radii to avoid overlap. In the Elk library, the muffin tin radii for Ni and As are the same, equal to 2.4. After optimization, the code reduces them to 2.205.} 
The stability requirement for self-consistency calculations was set to the total energy variation within 10$^{-5}$\,eV and for the Kohn–Sham potential $10^{-7}$\,eV.\\
(iii) Calculations were performed with the inclusion of the spin-orbit coupling and spin polarization effects, accounting for Eu$^{2+}$ moments in an incommensurate antiferromagnetic helical structure with a magnetic propagation vector $k =(0,0,0.92)$ determined by neutron diffraction experiments \cite{Jin2019}. 
\section{Magnetic Properties}\label{MagneticProperties} 
\begin{figure}[t] 
		\centering 
		\includegraphics[width=0.75\columnwidth]{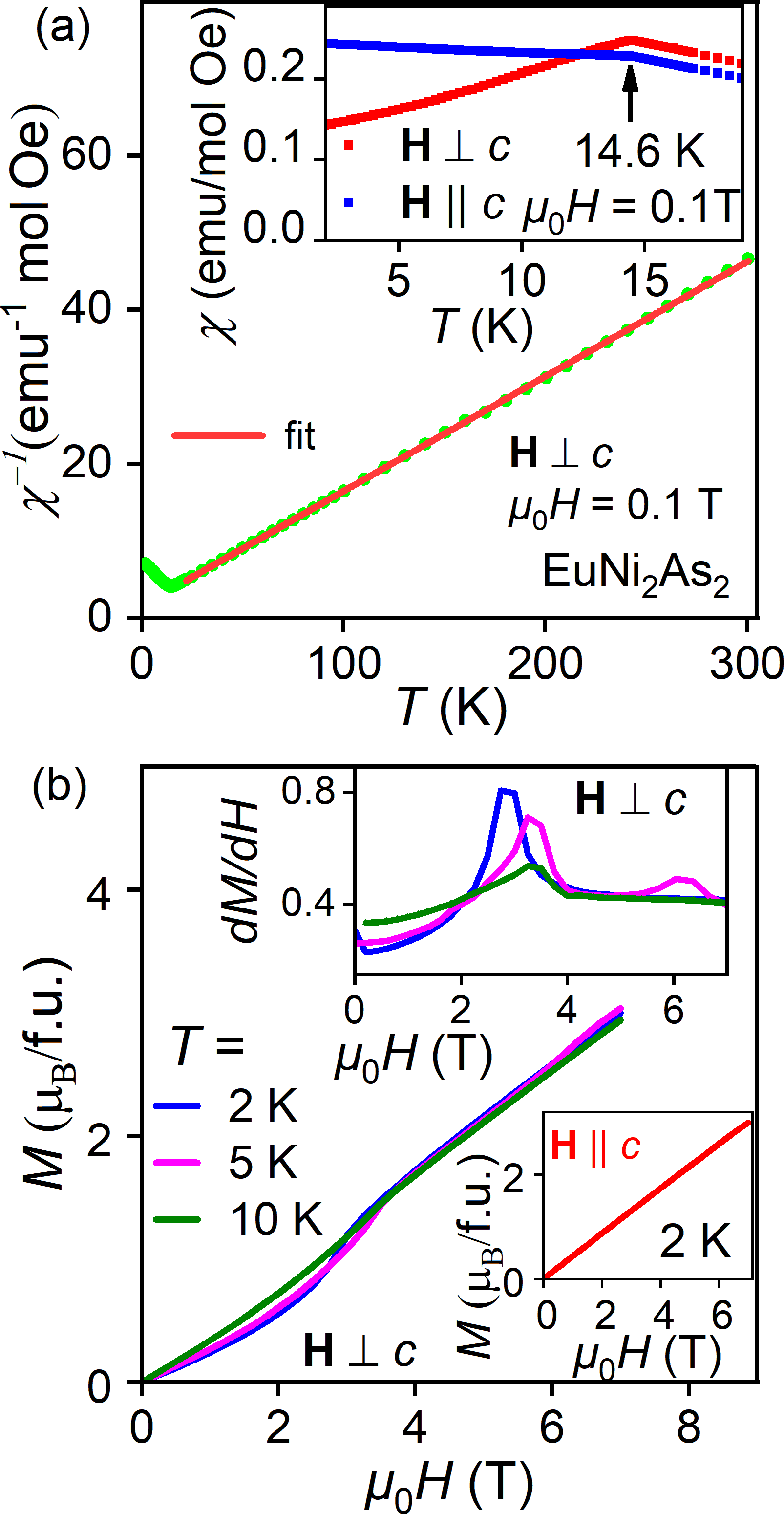} 
		\caption{(a) The temperature dependence of the inverse of susceptibility. The inset shows $\chi^{-1}(T)$ at $T<20$\,K, for both $\textbf{H}\perp c$ and $\textbf{H}\parallel c$. (b) The field dependence of the magnetization $M$ for $\textbf{H}\perp c$ and, in lower inset - for $\textbf{H}\parallel c$, the upper inset shows derivative $dM/dH$ for $\textbf{H}\perp c$, in units of $\mu_{B}/(\text{T f.u.)}$.}
		\label{Fig1Magn} 
\end{figure}

Temperature dependence of the inverse of the magnetic susceptibility, $\chi^{-1}(T)$, and magnetization as a function of magnetic field, ${M(H)}$ are shown in Fig.~\ref{Fig1Magn}. 

The Curie-Weiss law: $\chi^{-1}= (T-\theta_{\rm p})/C$
was fitted to the data collected in the magnetic field $\mu_{0}H=0.1$\,T, applied perpendicular to the $c$-axis, at temperatures between 40\,K and 300\,K, as shown in Fig.~\ref{Fig1Magn}(a). Here $C=\mu_{0}N\mu^{2}_{\rm eff}/3k_{\rm B}$ is the Curie parameter and $\theta_{\rm p}$ is the paramagnetic Curie-Weiss temperature. The parameters obtained from the fit are as follows: $\theta_{\rm p}=-11$\,K and the effective magnetic moment, $\mu_{\text{eff}}=7.33\,\mu_{\text{B}}$. The $\mu_{\text{eff}}$ is slightly lower than the theoretical value for Eu$^{2+}$ ($7.94\,\mu_{\text{B}}$), while $\theta_{\rm p}$ is close to the previously reported \cite{Sangeetha2019}. 
 $\chi(T)$ measured in $\mu_{0}H=0.1$\,T for two field directions: $\textbf{H}\parallel c$ and $\textbf{H}\perp c$-axis, confirms an AFM ordering below  $T_{\rm N}$=\,14.6\,K,
 consistent with previous reports \cite{Sangeetha2019, Jin2019} (see the inset of Fig.~\ref{Fig1Magn}(a)). 
 
$M(H)$ was also measured for both directions: $\textbf{H}\perp c$ and $\textbf{H}\parallel c$, as shown in Fig.~\ref{Fig1Magn}(b) and its lower inset, respectively. It attains $\approx3\,\mu_{\rm B}$/f.u. at 7\,T, for both field directions. Metamagnetic transitions (MMTs) are visible for $\textbf{H}\perp c$. The transition at $\approx3$\,T becomes gradually less pronounced upon increasing $T$ from 2 to 10\,K, but the second noticeable transition occurs at $T$=\,5\,K, just above $\mu_{\rm 0}H$=\,6\,T, clearly revealed by $dM/dH$ presented in the upper inset of Fig.~\ref{Fig1Magn}(b). On the other hand, $M(H)$ in $\textbf{H}\parallel c$ is linear, as shown in the lower inset, without noticeable trace of transitions. These observations are in accord with the report of Sangeetha et al. \cite{Sangeetha2019}. 

\section{Magnetotransport Properties} 
\begin{figure*}[] 
		\centering 
		\includegraphics[width=0.75\textwidth]{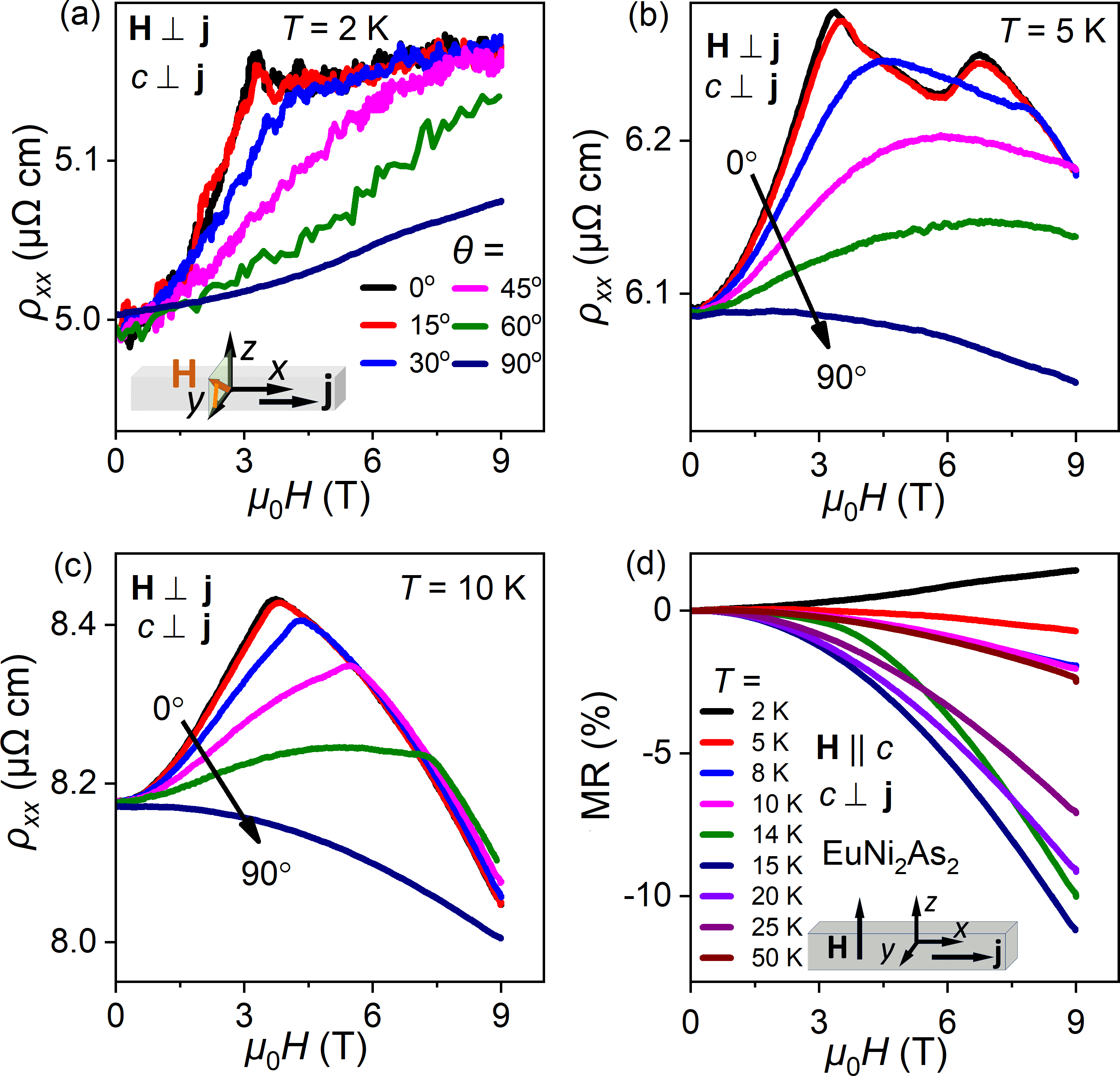}
		\caption{The magnetic field dependence of resistivity in different configurations for (a) $T$=\,2\,K, (b) 5\,K and (c) 10\,K. The color scheme for (b) and (c) is identical to (a). (d) The magnetoresistance for various temperatures above and below $T_{\rm N}$.}
		\label{Fig2MR} 
\end{figure*} 
The magnetic field dependence of resistivity, $\rho_{xx}$, of EuNi$_2$As$_2$ is shown in Fig.~\ref{Fig2MR} for various temperatures ranging from $T$\,=\,2 to 50\,K, in different configurations.
%(the geometry of each measurement is displayed in corresponding Figure, where $x, y$ and $z$ are along the crystallographic axes $a, b$ and $c$) with the magnetic field always applied in a direction transverse to the current. 
Figs.~\ref{Fig2MR}(a, b and c) show $\rho_{xx}(H)$ for field applied at various angles, at 2, 5 and 10\,K respectively. The experimental configuration was such that  $\textbf{H}$ remained perpendicular to the current, \textbf{j}, but was rotated from perpendicular to parallel to the $c$-axis, as schematically depicted in inset to Fig.~\ref{Fig2MR}(a). The $\rho_{xx}(H)$ increases in the weak magnetic fields for given temperatures with a hump close to $\mu_{0}H=$ 3\,T for 2\,K that shifts slightly toward higher field with increasing temperature, followed by a decline for 5 and 10\,K. Interestingly, for 5\,K, another hump arises in the magnetic field above 6\,T. These anomalies in $\rho_{xx}(H)$ align with the MMTs reflected in $M(H)$, for $\textbf{H} \perp c$, and vanish as \textbf{H} rotates toward $c$-axis. 

\begin{figure}[] 
		\centering 
		\includegraphics[width=0.75\columnwidth]{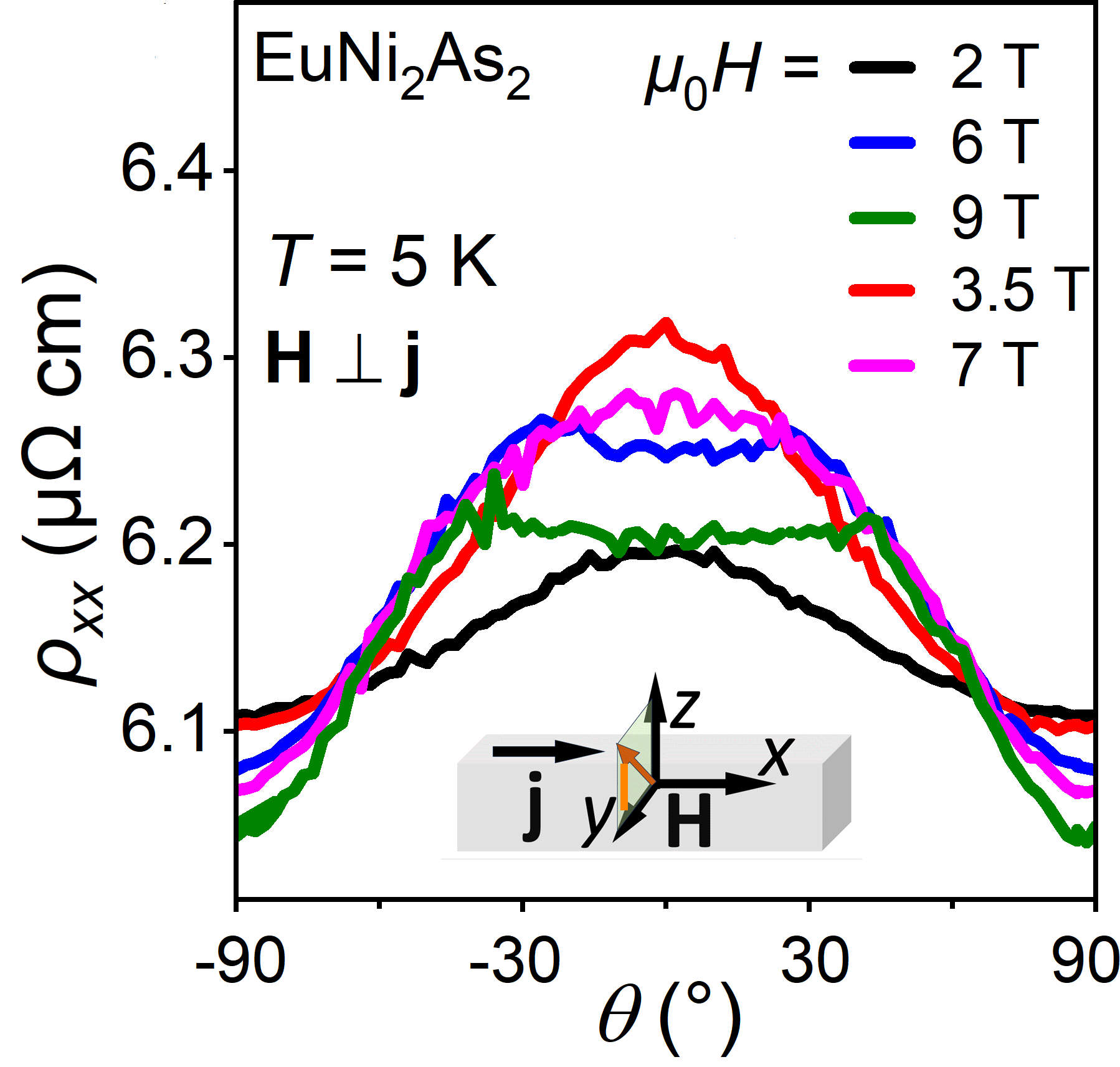}
		\caption{Field-angle dependence of longitudinal resistivity at 5\,K, for various magnetic fields.} 
		\label{fig-ang-rho} 
\end{figure}

The angular dependence of $\rho_{xx}$ at $T$=\,5\,K, for various magnetic fields is shown in Fig.~\ref{fig-ang-rho}. The geometry of measurement was similar to that shown in  Fig.~\ref{Fig2MR}(a), however, this time \textbf{H} was rotated continuously by angle $\theta$ towards $c$-axis, as shown in inset to Fig.~\ref{fig-ang-rho}, while the field strength and the temperature were fixed. Interestingly, $\rho_{xx}(T=$5\,K) at $|\theta|\approx65^{\text{o}}$ for all of the fields collapses to nearly identical value. This, means that the magnetoresistance is negligibly small at that angle or, in other words, a compensation occurs of the mechanisms inducing negative and positive magnetoresistance. 

To visualize the field variation of $\rho_{xx}$ for ($\textbf{H} \parallel c\perp {\mathbf j}$), we calculated the magnetoresistance using the relation: MR=$[\rho(H)/\rho(0)]-1$. For $T$=\,2\,K, MR is positive, reaching 1.5\% in $\mu_{0}H$=\,9\,T as shown in Fig.~\ref{Fig2MR}(d). For $T=$ 4\,K and above, it turns negative, and becomes the most pronounced ($-11.2\%$ for 9\,T) as 15\,K is reached, just above $T_{\rm N}$. It then starts to increase with increasing temperature, up to $-2.5$\% at 50\,K. 

The negative MR seems due to reduction of the spin-disorder scattering, a mechanism described by the de~Gennes and Friedel \cite{dGF_1958anomalies}. It associates the magnetic field dependence of paramagnet's resistivity with its magnetization as: $\rho_{xx}(H)\propto(1-M^2)$ \cite{duTdeL_2002}. We approximated $M(H)$ with the Brillouin function and achieved excellent fitting to the experimental $\rho_{xx}(H)$ data collected at 15\,K$\leq T\leq$50\,K, as shown in Fig.~\ref{dGF-fig}. Details of that model fitting and obtained parameters are presented in Supplemental Materials \cite{SupplMater}.

Several compounds have been reported that display the THE due to the spin chirality, emerging at the MMTs. The chiral spin structure can add Berry phase to the wavefunctions of conduction electrons, which leads to phenomena like THE \cite{Shang2021,Roychowdhury2023}.
However, for our samples of EuNi$_2$As$_2$, the measurements of Hall resistivity for the field direction, in which MMTs were observed are virtually impossible. This is because the single crystals grew very thin in $c$-direction and therefore we could not prepare the sample with the voltage contacts spanning along $c$-axis. 

The magnetic field dependence of the Hall resistivity, $\rho_{xy}$, for various temperatures from $T$=\,2 to 50\,K , in fields up to 9\,T, is displayed in Fig.~\ref{HallRes}. The slightly curvilinear $\rho_{xy}(H)$ is observed for $T$=\,2\,K and 4\,K, up to the highest applied magnetic field. Such an observation reveals the presence of more than one band crossing the Fermi level. For higher temperatures, namely $\geq$4\,K, linear fits of  $\rho_{xy}(H)=\mu_0H/(en)$, were performed to estimate the carrier concentration $n$ ($e $ is the elementary charge). A positive ordinary Hall coefficient $(en)^{-1}$ indicates that the holes are dominant charge carriers in EuNi$_2$As$_2$. The carrier density is of the order of $10^{22}\,$cm$^{-3}$, and changes moderately with increasing temperature as shown in the inset of Fig.~\ref{HallRes}. This reconfirms metallic nature of this compound, in accordance with its very small Hall resistivity. Unfortunately this impedes observation of THE, which usually bring out the topological $\rho_{xy}^{\rm T}$ being usually only a few percent of the total $\rho_{xy}$ \cite{Singh2024}. 
\begin{figure}[] 
		\centering 	
		\includegraphics[width=\columnwidth]{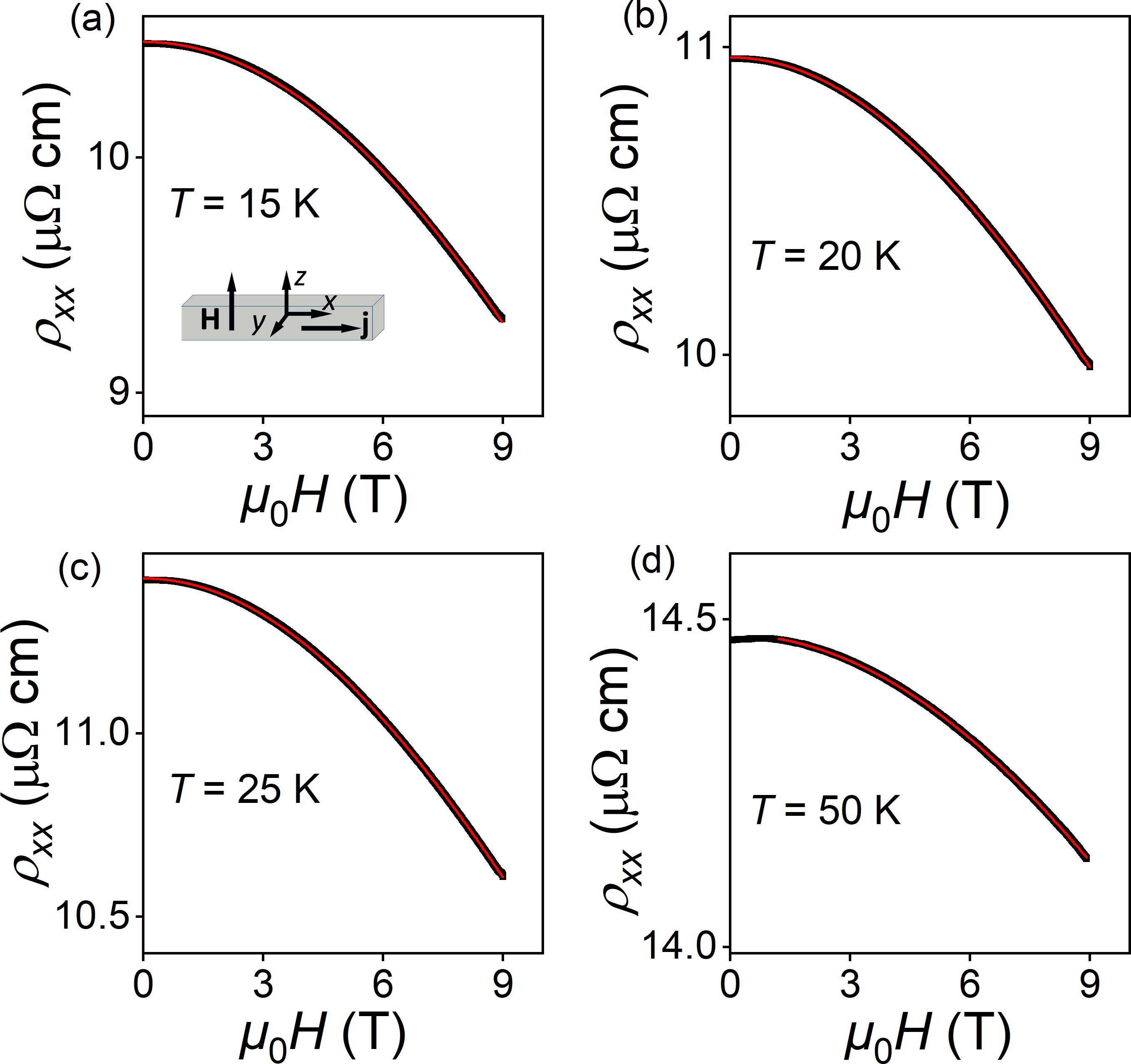}
		\caption{The magnetic field dependence of the longitudinal resistivity for (a) $T=$15\,K, (b) 20\,K, (c) 25\,K, and (d) 50\,K. The red solid lines show the fits of the de~Gennes-Friedel model, detailed explanation and fit parameters are presented in Supplemental Material \cite{SupplMater}).} 
		\label{dGF-fig} 
\end{figure} 
\section{Electronic structure}  
There are no spectroscopic data available for EuNi$_2$As$_2$, therefore instead of the commonly used value of Coulomb correlation potential, $U=$8\,eV for Eu$^{2+}$ \cite{Li2012, Paramanik2014, Jeevan2008}, we used several values of $U$ from 5 to 8\,eV at intervals of 0.5\,eV, including double counting correction in fully localized limit (FLL) \cite{liechtenstein1995density}. It occurred, $U=$5\,eV was the most suitable value with the least total energy among the rest, with a modest difference in the electronic band calculations. Hence, all the subsequent calculations were performed for this value of $U$. The characteristic features of band structure for the considered values of $U$ are presented in Tab.~\ref{Table-DFT}.

\begin{figure}[] 
		\centering 
		\includegraphics[width=0.8\columnwidth]{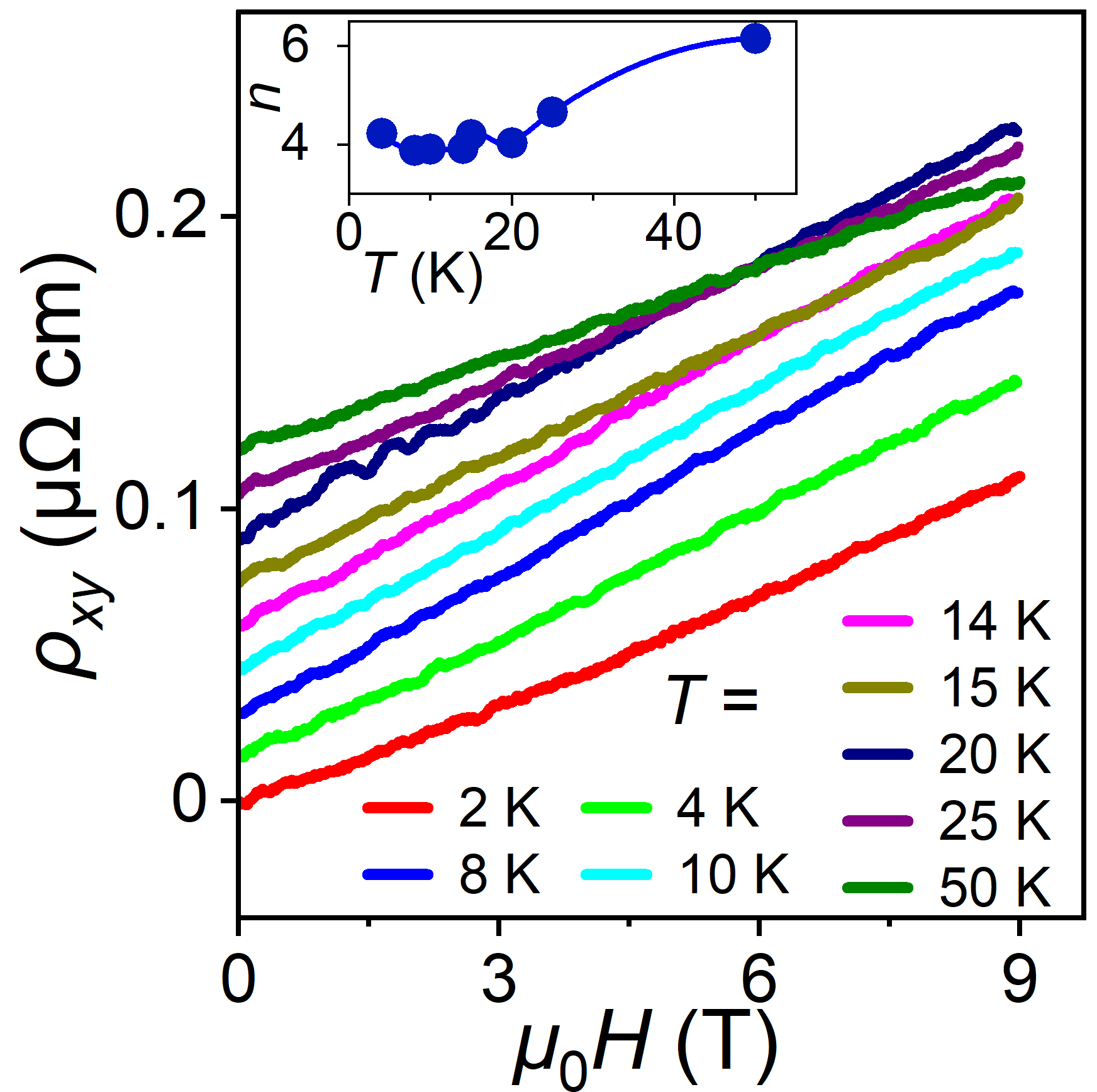}
		\caption{Magnetic field dependence of Hall resistivity for  temperatures ranging from $T$=\,2 to 50\,K (data are offset for clarity's sake). Inset shows charge carrier concentration, $n$, in 10$^{22}$\,cm$^{-3}$ units.} 
		\label{HallRes} 
\end{figure} 
  
\begin{figure*}[] 
		\centering 
	\includegraphics[width=0.6\textwidth]{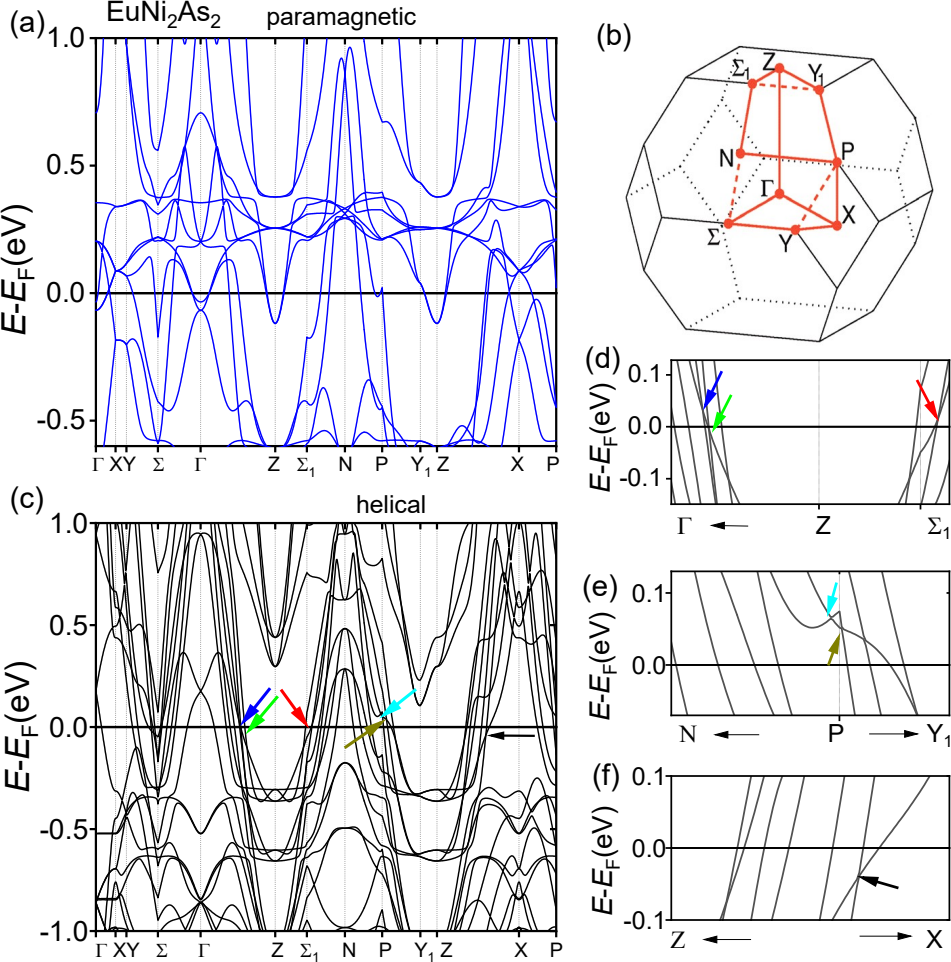}
		\caption{The electronic band structure of EuNi$_2$As$_2$, calculated with Coulomb correlation potential $U$=\,5\,eV and Hund's coupling $J$=\,0.8\,eV, (a) for the paramagnetic state and (c) for the helical AFM state. (b) The sketch of the Brillouin zone for the $I4/mmm$ space group, with high symmetry lines shown in red. (d-f) Magnified regions of the band structure shown in (c) near the Fermi level ($E$=0).}
		\label{DFTafm} 
\end{figure*}

The electronic band structure for $U$=\,5\,eV in the paramagnetic state is shown in Fig.~\ref{DFTafm}(a), with a finite number of hole and electronic pockets. Fig.~\ref{DFTafm}(b) displays the Brillouin zone sketch of EuNi$_{2}$As$_{2}$. The electronic band structure for the helical AFM state is shown in Fig.~\ref{DFTafm}(c). In the paramagnetic region four 2-fold degenerate bands cross the Fermi surface. This degeneracy is lifted in the AFM state, where 8 bands cross the Fermi surface. 
   Furthermore, the band structures with various $U$ values for the helical AFM state were almost identical (comparison not shown here for simplicity's sake) near the Fermi level, indicating the dominant metallic behavior. The number of electron and hole pockets contributes to the total conductivity through the Fermi level, $E_{\rm F}$. We observe multiple Dirac-like cones in the helical AFM phase in the close vicinity of $E_{\rm F}$ marked with arrows in Figs.~\ref{DFTafm}(d-f). However, the contribution of topologically nontrivial states is not detected in our transport measurements, most likely overwhelmed by the contributions from trivial bands. 
   
\begin{figure*}[] 
		\centering 
	\includegraphics[width=0.6\textwidth]{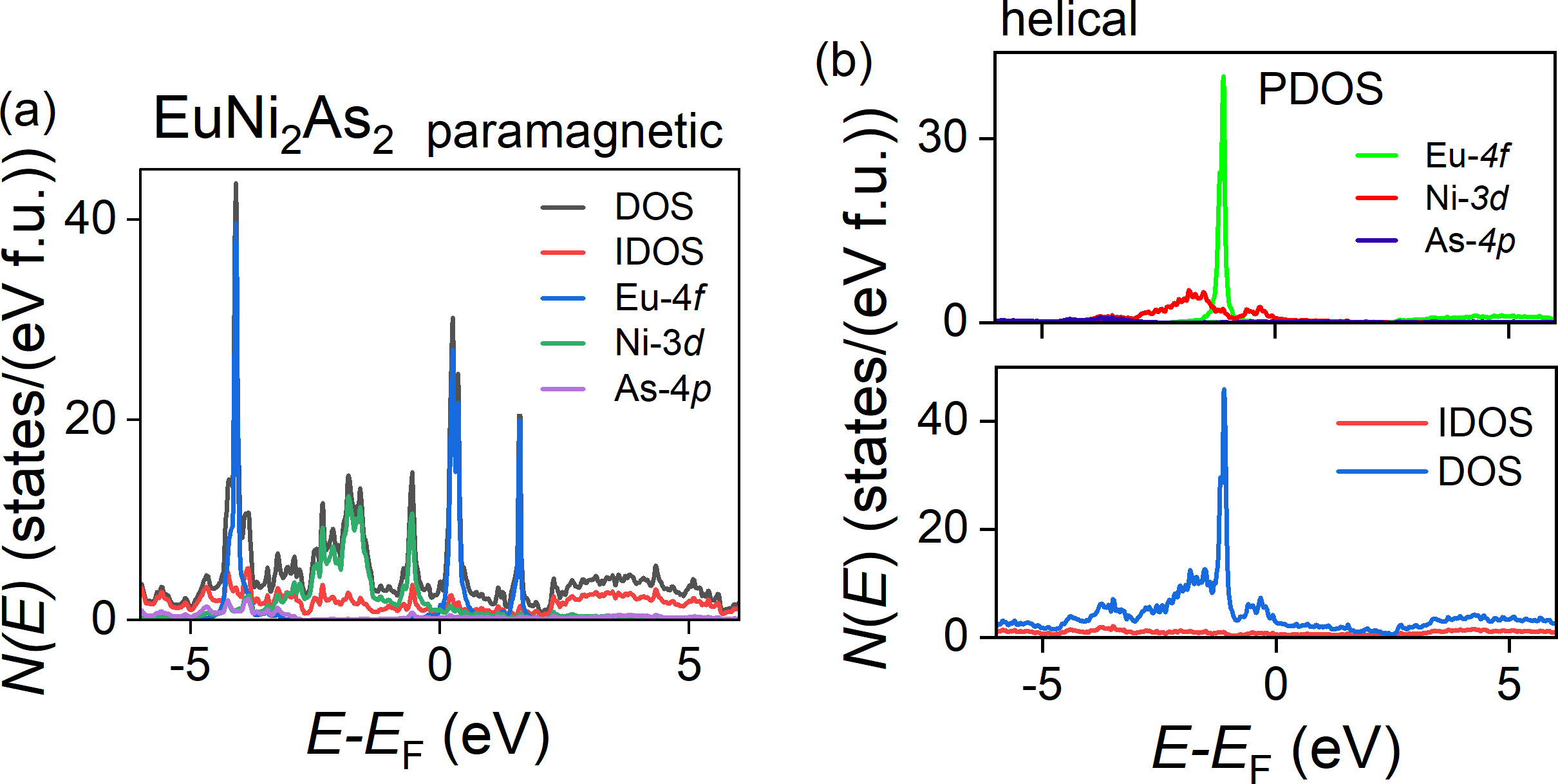}
		\caption{Interstitial (IDOS), and total (DOS) density of states: (a) for the paramagnetic and (b) for the helical AFM phase. The  PDOS shown in (b) are sums of the identical spin-up and spin-down contributions.} 
		\label{FigDOS} 
\end{figure*}

  To gain a deeper understanding of the electronic band structure, we also calculated the density of states (DOS). The partial (PDOS), interstitial (IDOS), and total (DOS) are illustrated in Fig.~\ref{FigDOS} for paramagnetic and ordered phases. The Eu-4\textit{f} level lies above $E_{\rm F}$ in the paramagnetic state and shifts below $E_{\rm F}$ in the ordered phase. The 4\textit{f} bands exhibit a minor contribution at $E_F$, while spiking well below the Fermi level, thus resulting in weak hybridization with Ni-3\textit{d} states. 
  The FLL correction within the LSDA+$U$ method clearly results in a greater degree of \textit{f}-electrons' localization than the GGA method calculations reported by Tran et al. \cite{tran2025first}. Whereas the Ni-3\textit{d} band is the major contributor to the total conductivity at the Fermi surface, the As-4\textit{p} band negligibly contributes to the total conductivity, as shown in Fig.~\ref{FigDOS}(b). 
  The IDOS and DOS are depicted in the lower panel of Fig.~\ref{FigDOS}(a). Interestingly, 20\,$\%$ of DOS value at the Fermi level is coming from the contribution of IDOS. DOS($E_{\rm F}$) in the paramagnetic and helical AFM phases changes from 3.3 to 6.13\,states/(eV\,f.u.), respectively, which reflects the modest change in carrier concentrations between the two regions.

The spin-polarized data of all calculations support the magnetic order and show that the essential contribution to the magnetic moment should come from Eu-4\textit{f} ions. The estimated total magnetic moments presented in Tab.~\ref{Table-DFT} are comparable to the 
experimental value of 6.75(6) $\mu_{\rm B}$/Eu reported earlier \cite{Jin2019}. 
Using the screened Coulomb interaction in the form of a Yukawa potential proposed by Norman et al. \cite{Norman}, by choosing $U$ and $J$ individually, we obtained the screening length values for 4\textit{f} shells. (Tab.~\ref{Table-DFT}). The small values of the screening length of 4\textit{f} orbitals of Eu$^{2+}$ are due to the fact that they are deeply shielded by the filled 5\textit{s} and 5\textit{p} orbitals, which in turn shield the 4\textit{f} electrons from their surrounding environment.
We observe that at the fixed $J$=\,0.8\,eV, the increasing Hubbard $U$ term from 5 to 8\,eV results in a slight growth in the magnetic moments from 6.98 to 7.041\,$\mu_{\rm B}$, while the screening length decreases.
We observe the LSDA+$U$ moments in the limit of screening length. We note that by gradually decreasing the $\lambda$ parameter, one increases the localization of the 4\textit{f} electrons. The similar behavior was also found in the case of 5\textit{f} shells in U, Np, Pu-based actinide compounds \cite{bultmark2009multipole}.
\section{Conclusion and summary}
We grew and characterized single crystals of EuNi$_{2}$As$_{2}$. We remeasured their magnetic properties, which reaffirmed that the magnetic order takes place below $T_{\rm N}$=\,14.5\,K and the effective magnetic moment is 7.33\,$\mu_{\rm B}$.  We measured and analyzed the transverse magnetoresistance for different orientations of the applied magnetic field. The MR for $\textbf{H}\parallel c$-axis is about few percent and negative at low temperatures, which for $T>T_{\rm N}$ we ascribe to the de~Gennes-Friedel mechanism. The anomalies in $\rho_{xx}(H)$ for $\textbf{H}\perp c$-axis corresponds well to MMTs, which are observed in magnetization for the same direction of the applied magnetic field. Occurrence of such anomalies in MR are usually ascribed to the Berry curvature in real space due to the spin chirality, and they are accompanied by THE. Unfortunately, the configuration of Hall measurement with $\textbf{H}\perp c$ in inaccessible for our very thin in $c$-direction samples. Only cutting crystals with focused ion beam could allow preparation of the sample in such configuration, but nevertheless we doubt, that THE could be observed in our metallic compound. 
\begin{table*}[]
\renewcommand{\arraystretch}{1.3}
\begin{center}
\caption{Relative energies of the system ($E - E_0$), total densities of states at $E_{\rm F}$, 
    screened length and magnetic moments of Eu$^{2+}$ ions calculated with the LSDA+U approach. 
    $E_0$ denotes the energy calculated with $U=$\,0.} \label{Table-DFT}
\begin{tabular}{l . . . . . . . . .} 
\hline
\textit{U} (eV) & 5& 5.5 & 6 & 6.5 & 7 & 7.5 & 8\\ 
\hline\hline
$(E-E_0)$ (me\,V)&-15.73& -15.58 &-15.44 & -15.31&-15.19  &-15.07  &-14.96  \\ 
$N(E_{F})$(state/(eV\,f.u.))   &2.708&2.722 & 2.802&2.73&2.756 &2.791 &2.803 \\
screening length ($\lambda$)&2.37& 2.19 & 2.03 & 1.89 & 1.76 & 1.65 & 1.55 \\ 
magn. moment ($\mu_{B}$/f.u.)	&6.988& 6.999 & 7.009 & 7.018 &7.026& 7.034 &7.041 \\ 
\hline
\end{tabular}
\end{center}
\end{table*}  

Further, by analyzing the Hall resistivity  measured in $\textbf{H}\parallel c$, we found that multiple bands contribute to the conductivity at low temperatures, whereas the single band model fits very well at higher temperatures, resulting in the carrier concentration of the order of $10^{22}\mathrm{cm}^{-3}$. Probably that is why, unlike in the EuCuAs \cite{Roychowdhury2023}, the topological Hall effect was not observed in EuNi$_{2}$As$_{2}$. The component of THE due to the Berry curvature in momentum-space brought by plausible Dirac/Weyl points is obscured by the large concentration of carriers from topologically trivial bands, making the overall Hall resistivity very low. Survey of literature showed that usually the $\rho_{xy}$ due to THE is a few percent of the total Hall resistivity \cite{Singh2024}. Therefore, for EuNi$_{2}$As$_{2}$ it might be expected about $10^{-7}\Omega\,\text{cm}$, which is beyond the sensitivity of our measurements. 

We also revisited the specific heat with employing the Debye-Einstein model, and found characteristic parameters, which are given in Supplemental Materials \cite{SupplMater}. Further, we incorporated the \textit{ab-initio} calculations, which provided deeper insight into electronic band structure of the compound. The changes in the density of states, magnetic moments, and screening lengths of Eu-4\textit{f} orbitals were discussed in terms of the effects of Hubbard corrections. We found that interstitial DOS has a significant role in total DOS. Despite strong shift in position of Eu-4$f$ band occurring with the magnetic ordering, the density of states at the Fermi level decreases only by a factor smaller than two.
\\\\
\textbf{ACKNOWLEDGMENT}\\
This study was supported by the National Science Centre (Poland) under grant 2021/41/B/ST3/01141.

	\bibliography{aaa-myrefNi} 
	%\printbibliography
\clearpage	
\onecolumngrid
 \begin{center}
  \textbf{\Large Supplemental Material}\\[.2cm]
  \textbf{\large  Magnetotransport and electronic band structure of EuNi$_2$As$_2$ antiferromagnet}\\[.2cm]
Faheem Gul, Mane Sahakiyan, Orest Pavlosiuk and Piotr Wiśniewski\\  %, Dariusz Kaczorowski

  {\itshape $^{1}$Institute of Low Temperature and Structure Research, Polish Academy of Sciences, Wrocław, Poland\\      
  }
(Dated: \today)
\\[1cm]
\end{center}
\setcounter{equation}{0}
\renewcommand{\theequation}{Eq. S\arabic{equation}}
\setcounter{figure}{0}
\renewcommand{\thefigure}{S\arabic{figure}}
\setcounter{section}{0}
\renewcommand{\thesection}{S\arabic{section}}
\setcounter{table}{0}
\renewcommand{\thetable}{S\arabic{table}}
\setcounter{page}{1}
%%%%%%%%%%%%%%%%%%%%%%%%%%%%%%%%%%%
\begin{doublespace}
\section{Crystal Growth and Characterization} 
\begin{figure}[H] 
		\centering 
		\includegraphics[width=0.7\textwidth]{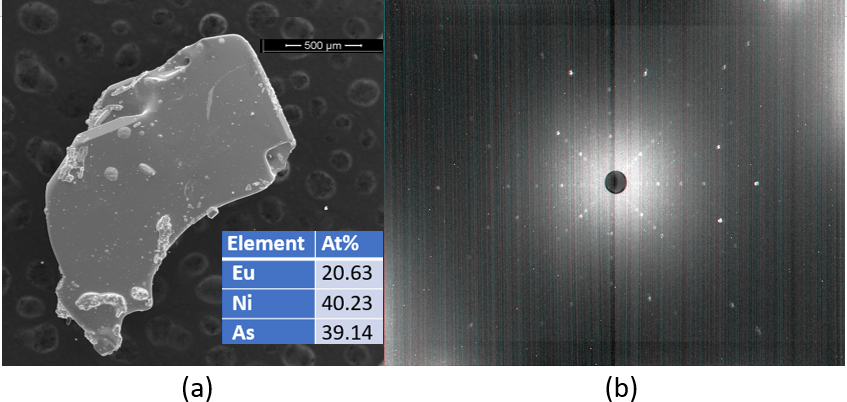}
		\caption{(a) SEM image of a single crystal of EuNi$_2$As$_2$. The inset shows elemental composition from EDS analysis. (b) Laue diffraction pattern taken with incident x-ray beam close to the [001] crystallographic direction.} 
		\label{Laue_EDS} 
\end{figure}

Single crystals of EuNi$_2$As$_2$ were grown using the Bi-flux technique.  Eu, Ni, As, and Bi were taken in 1:2:2:10 molar ratios. The elements were put into an alumina crucible, vacuumed and sealed in a quartz ampule. The ampule was heated to 530\,°C at the rate of 25\,°C/h, followed by a 24 hours dwell. Next, temperature was raised to 1050\,°C and kept for 24 hours for complete homogenization. Then the ampule was cooled  to 500\,°C at 2\,°C/h rate. Finally, platelet-like crystals were separated from flux with a centrifuge. Figs.~\ref{Laue_EDS}(a) and \ref{Laue_EDS}(b) show scanning electron microscope (SEM) image, elemental composition, and a Laue diffraction pattern of a single crystal of EuNi$_2$As$_2$, respectively.
	
\section{Heat capacity}
\begin{figure}[] 
		\centering 
		\includegraphics[width=0.8\textwidth]{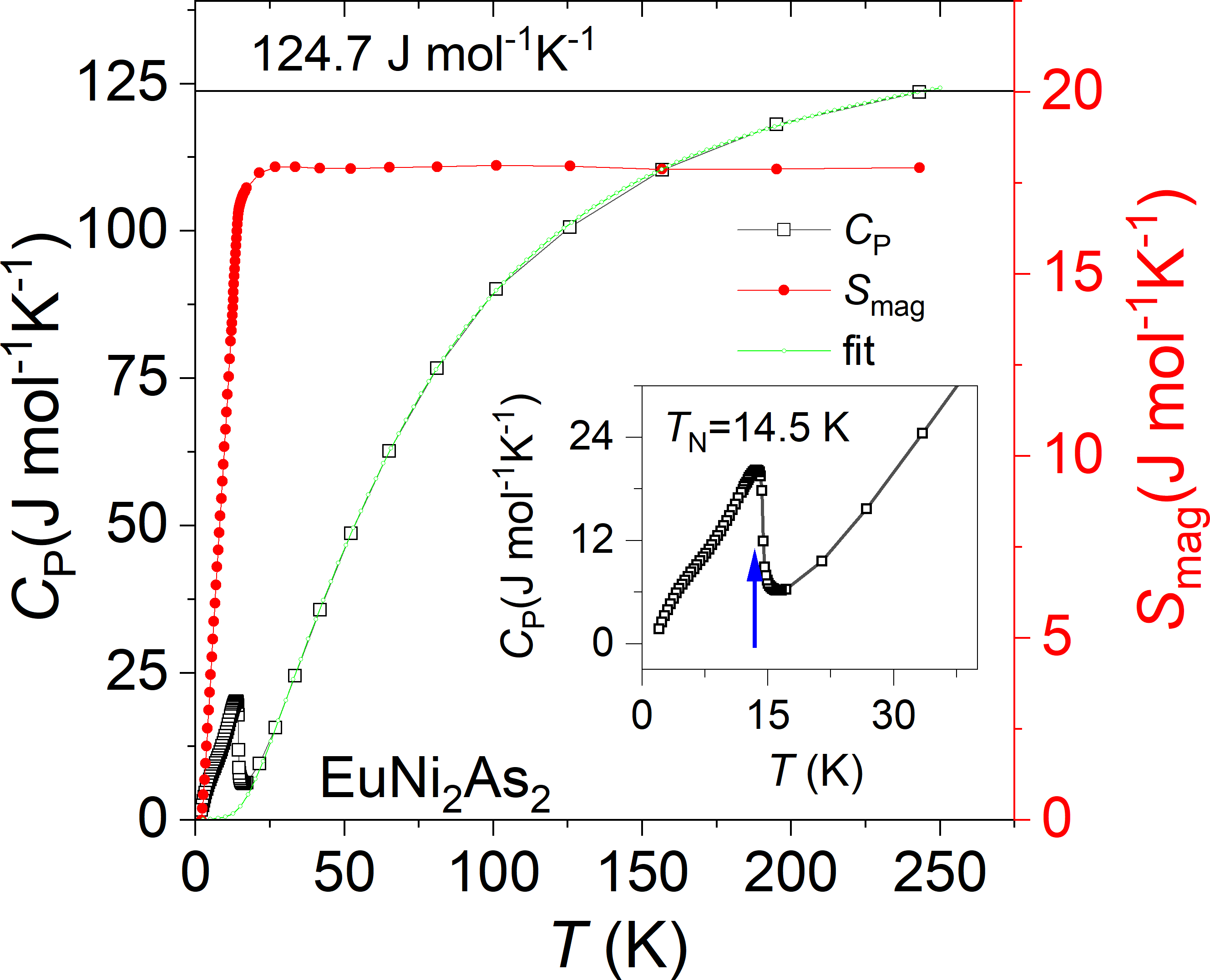}
		\caption{The temperature dependence of the specific heat ($C_{\rm p}$ - left axis) and the magnetic entropy ($S_{\rm{mag}}$ - right axis) of EuNi$_2$As$_2$. The green solid line shows the fit with Debye-Einstein model. The inset shows anomaly of $C_{\rm p}$ at $T_{\rm N}$=\,14.5\,K.} 
		\label{HCNickel} 
\end{figure}

The specific heat ($C_{\rm p}$) of EuNi$_2$As$_2$ was measured in the temperature range 2 $\le \textit{T} <$ 300\,K, shown in Fig.~\ref{HCNickel}. The $\lambda$-shaped anomaly associated with AFM transition is visible near \textit{T} = 14.6\,K and shown enlarged in the inset of Fig.~\ref{HCNickel}. $C_{\rm p}$ attains the classical Dulong-Petit limit just close to the $T$\,= 250\,K, that is $3n\mathcal{R}$ =\,124.7\,J\,mol$^{-1}$K$^{-1}$, where $n$ is the number of atoms in the primitive unit cell and $\mathcal{R}$ is the universal gas constant.
To estimate the magnetic contribution to specific heat, $C_{\text{mag}}$, from $C_{\rm p}$, the Debye-Einstein model was used:
\begin{equation}
    C_{\rm p} =  C_{\text{mag}} + C_{\text{e}}+ p\,C_{\text{E}}+ (1-p)\,C_{\text{D}},
 \end{equation}
 where $C_{\text{e}}$ is the electronic contribution,  $C_{\text{E}}$ and $C_{\text{D}}$ are the Einstein and Debye terms, respectively, and $p$ describes the relative weight of the Einstein and Debye contribution to the $C_{\rm p}$, and $C_{\text{e}}=\gamma T$, where $\gamma$ is the Sommerfeld coefficient. The Debye and Einstein terms account for contributions from the acoustic and optical phonons, respectively:
\begin{equation*}
     C_{\text{D}}=9n\mathcal{R}\left(\frac{T}{\theta_{\text{D}}}\right)^3\int_{0}^{{\theta_{\text{D}}}/{T}}\frac{ x^{4} e^{x}}{(e^{x}-1)^{2}}{\rm d}x,~~{\rm and}~~ 
    C_{\text{E}}=3n\mathcal{R} \left(\frac{\theta_{\text{E}}}{T}\right)^2
    \frac{e^{\theta_{\text{E}}/T}}{(e^{\theta_{\text{E}}/T}-1)^2}.
 \end{equation*}
$\theta_{\rm D}$ and $\theta_{\rm E}$ denote characteristic Debye and Einstein temperatures, respectively. The fit (solid green line) shown in Fig.~\ref{HCNickel} yielded the following parameters:  $\theta_{\rm D}$=\,364(9)\,K, $\theta_{\rm E}$=\,119(3)\,K, $\gamma$=\,31(1)\,mJ\,mol$^{-1}$K$^{-2}$ and $p$=\,0.40(3). The magnetic contribution $C_{\rm mag}$ was used to estimate the magnetic entropy as 
$S_{\rm mag}$=\,$\int_{2\rm{K}}^{T}(C_{\text{mag}}/{T'})\,{\rm d}T'$.
 At $T=29$\,K, $S_{\rm mag}$ attains the maximum value of 17.9\,J\,mol$^{-1}$K$^{-1}$, only slightly different from the theoretical value for the Eu$^{2+}$ ion, calculated as $\mathcal{R}\,{\rm ln}(2s+1)$=\,17.3\,J\,mol$^{-1}$K$^{-1}$ with $s=7/2$, shown on the right axis of Fig.~\ref{HCNickel}. 
 In the previous report only the Debye model was employed, and yielded $\theta_{\rm D}$=\,280\,K and significantly lower $S_{\rm mag}$=\,15.38\,J\,mol$^{-1}$K$^{-1}$ \cite{Sangeetha2019}. 

Furthermore, we estimated the density of states at the Fermi level, $N(E_{\text{F}})$, using the $\gamma$ value and the formula: 
$N(E_{\rm F})= {3\gamma }/({2\pi^{2}k_{\rm B} \mathcal{N}})$,
 where $k_{\text{B}}$ denotes the Boltzmann constant and $\mathcal{N}$ is the Avogadro number. The obtained value is $N(E_{\text{F}})$ $\approx$ 14\,state/(eV\,f.u.), which corresponds to the metallic nature of the EuNi$_2$As$_2$. Moreover, the Fermi energy can be written as, $E_{\rm F}=3n/{2\rm DOS}(E_{\rm F})$, where $n$, is the carrier concentration. For $T=$10 and 50\,K, $n$ is $3.9\times10^{22}\rm cm^{-3}$ and $6.1\times10^{22}\rm cm^{-3}$, respectively. The resulting $E_{\rm F}$ changes from 0.7\,eV at 10\,K to 1.1\,eV at 50\,K. \vspace{-0.7cm}
\section{Resistivities and de Gennes-Friedel modeling of magnetoresistance.}
\begin{figure}[H] 
		\centering 		\includegraphics[width=0.8\textwidth]{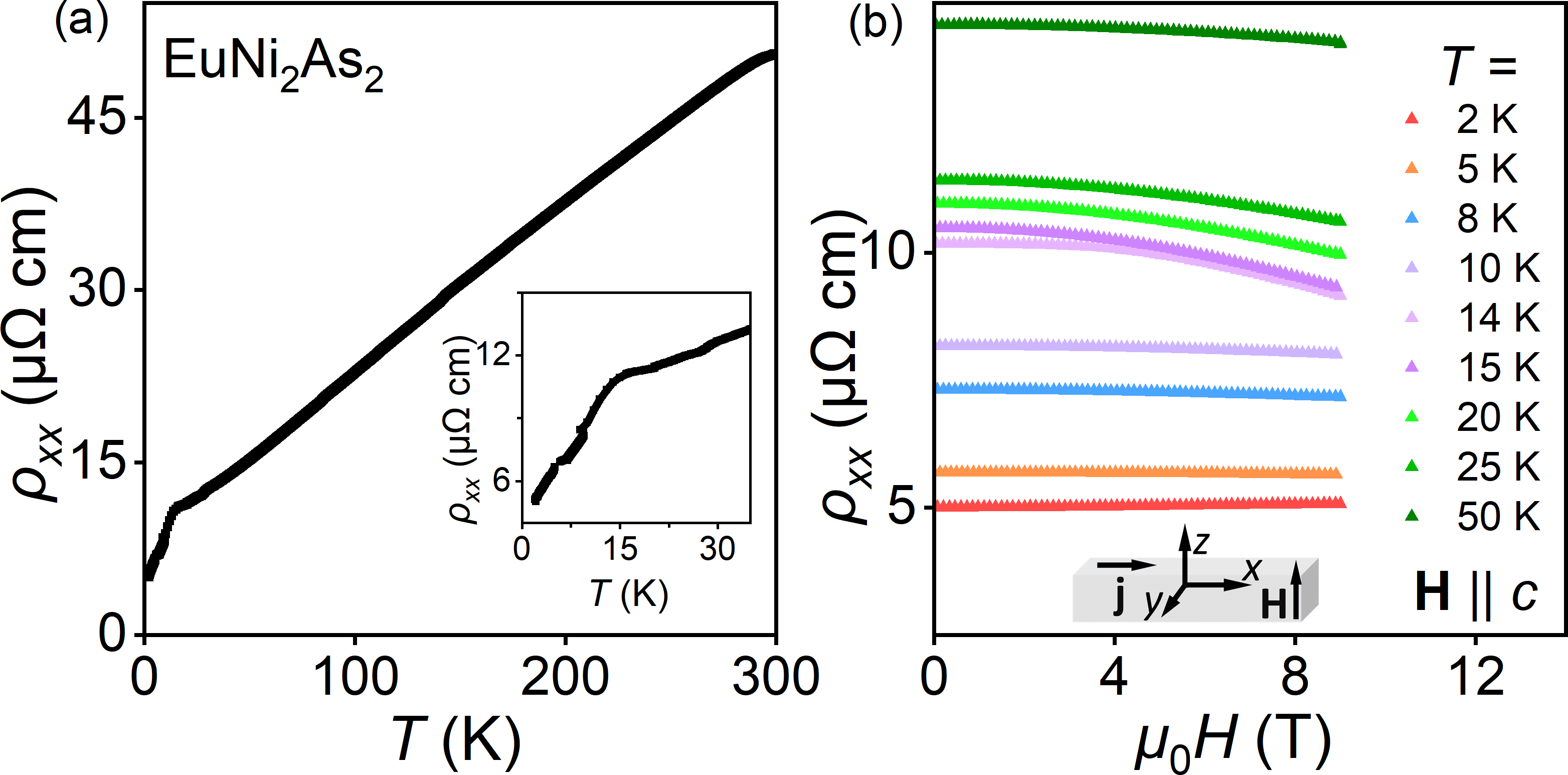}
		\caption{(a) The temperature and (b) magnetic field dependence of the $\rho_{xx}$.} 
		\label{resistivitySm} 
\end{figure}

  The temperature-dependent resistivity ($\rho_{xx}(T)$) is shown in Fig.~\ref{resistivitySm}(a), with its inset showing the data in low-temperature region. The residual resistivity ratio RRR= $\rho$(25\,K)/ $\rho$(300\,K)$\approx$ 4.1 reflects the high quality of synthesized single crystals. The field-dependent resistivity ($\rho_{xx}(H)$) is shown in Fig.~\ref{resistivitySm}(b) for various temperatures.
The suppression of spin-disorder scattering as spins get gradually aligned in the applied magnetic field, leads to negative magnetoresistivity. The mechanism was proposed by the de~Gennes and Friedel \cite{dGF_1958anomalies}, who proposed that the magnetoresistance reflects the field variations in the correlation function between spins, $\langle\mathbf{S}_i\cdot \mathbf{S}_j\rangle$, through the relation $\rho_{xx}\propto [S(S+1)-\langle\mathbf{S}_i\cdot \mathbf{S}_j\rangle]$. In the paramagnetic state it is related to the magnetization: $\rho_{xx}(H)\propto [1-M^2(H)]$ \cite{duTdeL_2002}. For magnetic moments of Eu$^{2+}$ we can very well approximate $M(H)$ with Brillouin function and get: 
\begin{equation}\label{dGF_eq}
    \rho_{xx}(T, H) = a [1-\big\{m_{T}\,\text{tanh}(H/t)\big\}^{2}]
\end{equation}
where $a$ denotes the $\rho_{xx}$ at zero field, $t$ determines the saturation field and $m_{T}$ scales the magnetization (which corresponds to the term in curly brackets). 
This model does not fit in the helical AFM region, where strong antiferromagnetic exchange interactions enter the correlation function, and $\langle\mathbf{S}_i\cdot \mathbf{S}_j\rangle/(S(S+1))$ no longer corresponds to $M^2$. The fits of Eq.~S2 for various temperatures in the paramagnetic region are shown in Fig.~\ref{dGF-fig} of the main text, and the fit parameters are collected in Table~\ref{dGF_tableSM}.  

\renewcommand{\arraystretch}{1.5}
\begin{table}[!h]
\begin{center}
\caption{Parameters obtained from fitting of the dGF model to $\rho_{xx}(H)$ isotherms above the Néel temperature.}
 \label{dGF_tableSM}\vspace{0.3cm}
\begin{tabular}{l l . l} 
\hline
$T$\,(K)~~&$a~(\mu\Omega\,\text{cm})$ &\multicolumn{1}{c}{$m_{\rm T}$}&$t~({\rm T})$ \\ 
\hline\hline
15& 10.4915(2)   & 1.18(1) &30.6(3) \\ 
20  & 10.96672(9) & 0.7302(1)& 20.38(5) \\
25	& 11.42224(8) & 0.689(2) & 22.01(7) \\ 
50  & 14.47652(3) & 0.547(2) &31.0(1)\\ 
\hline
\end{tabular}
\end{center}
\end{table}
\end{doublespace}
\end{document}